# Status of Hybrid Target R&D at KEK LINAC

Tohru Takahashi [*]

Graduate School of Advanced Sciences of Matter, Hiroshima University
1-3-1 Kagamiyama Higashi Hiroshima, 739-8530, Japan

A hybrid positron source, consisting of a crystal as electron to photon converter and a amorphous target, is a candidate to relax heat load problem for high intensity positron sources. In this article we report status of experimental study using the 8 GeV electron beam at KEKB LINAC.

## 1 Introduction

Future linear collider projects such as the International Linear Collider (ILC) and the Compact Linear Collider (CLIC) need intense positron beams to provide high luminosity for experiments [1,2]. The ILC requires about 3000 bunches in 1 ms and each bunch has $10^{10}$ positrons. The CLIC design assumes 300 bunches in 150 ns and they repeat every 20 ms. Positrons are usually produced by injecting driving electron beams in metal targets. However, as required intensity of the positron is getting higher, heat deposit in the metal targets increase and eventually destroys the targets. In addition, if one needs positron in a short time period, shock waves produced by driving electrons beam could damage the target. A way out of this problem is to impinge photons to relatively thin targets. However, it requires large number of photons of their energy higher than O(10MeV). The ILC adopts the undulator scheme for this purpose, where high energy electron beam of more than 150 GeV are fed into more than 100 m long undulator. This scheme is under development as the baseline design of the ILC. Recently, new scheme called "hybrid target" was proposed by authors to produce photons with relatively low energy electrons (see Figure 1)[3]. In the idea, a few to 10 GeV electrons are impinged into a thin crystal target. When the electron is impinged in parallel to the crystal axis of the target, the electrons pass through channels of crystal axes and radiate low energy photons. If we combine thick amorphous target as a positron generator downstream of the crystal, the crystal-amorphous target system could be a new source for intense positron beams.

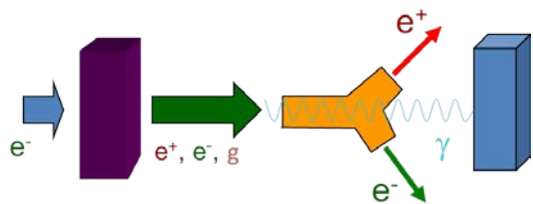

Figure 1. Schematic of hybrid positron sources

We performed experiment at KEKB LINAC to investigate performance of the hybrid source as a positron generator as well as to see technical feasibility in real accelerator environment. In this article, we report status of the experimental study at the KEKB LINAC.

## 2 Experimental Set Up

Experiments were performed at the beam switch yard of KEKB LINAC at High Energy Accelerator Research Organization (KEK), Japan. The energy of the electron beam is 8 GeV and typical bunch charge during the experiments was about 1.5 nC. The repetition rate of the bunch can be varied from 1 to 25 Hz.

---

[*] Collaborators: R. Chehab (IPNL), A.Variola (LAL), V. Strakhovenko (BINP), L. Rinolfi(CERN), O. Dadoun (LAL), T. Kamitani, T. Suwada, T. Omori, J. Urakawa, K. Furukawa, K. Umemori, M. Satoh, T. Sugimura(KEK) , S. Kawada, T. Akagi, Y. Uesugi (Hiroshima)

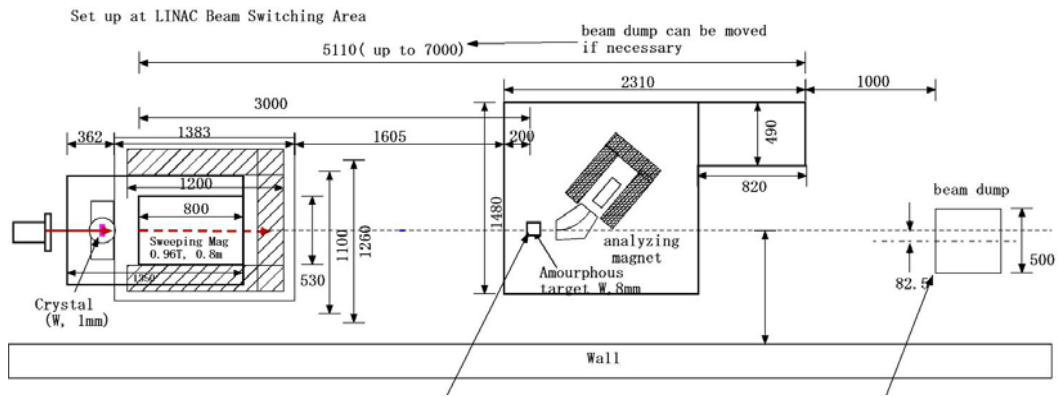

Figure 2. Experimental set up at the beam switch yard in KEKB LINAC

The 8 GeV beam was injected to the tungsten crystal mounted on the rotational table. After the crystal, a magnet was placed to sweep out charged particles coming out from the crystal. The magnet was surrounded by lead bricks of 20 cm thick to stop all charged particles including remnant of 8 GeV electrons. The shields have 20 mm ϕ on the beam line so that photons from the crystal were sent to the amorphous target placed about 3 m downstream of the crystal. At the amorphous targets area, 5 tungsten targets of 1.75 mm, 3.5 mm, 5.25 mm, 8mm, and 18 mm thick were placed on a remotely controlled stage. The thickness of 1.75 mm, 3.5 mm, and 5.25 mm correspond to 0.5, 1, and 1.5 radiation length of tungsten while 8 mm and 18 mm targets are chosen to provide maximum positron yields for the hybrid configuration and direct 8 GeV electron injection, respectively. On the back of the 8 mm and 18 mm thick target, thermocouples are attached to measure the temperature of the target due to energy deposit caused by the electro-magnetic shower in the target. After the amorphous target, positron momentum was analyzed by a dipole magnet and positrons of selected momentum were sent to a Cherenkov detector to count positron yields. A beam pipe was installed from the analyzing magnet to the detector and vacuum level inside of the pipe was about $10^{-3}$ Pa to reduce multiple scattering of low momentum positrons. A set of collimators were installed to improve momentum resolution of the system and to reduce background of the detector.

# 3 Results

## 3.1 Positron yield

Figure 3 shows measured positron yields from the 8 mm tungsten by changing <111> axis with respect to the electron beam. The analyzing magnet was set to choose 20 MeV positrons from the target. A peak of positron yields was clearly observed at specific angle where crystal axis and the electron beam are in parallel each other. The observed enhancement of positron yield at the aligned to off-aligned angle was about 3 in this configuration. It should be noted that observed width of the peak is wider than those expected from critical angle of the channeling, which is about 0.5 mr for 8 GeV electrons. It is because electrons slightly off angle from the channeling condition are still affected by axial potential of the crystal and radiate photons more than electrons of completely off axis.

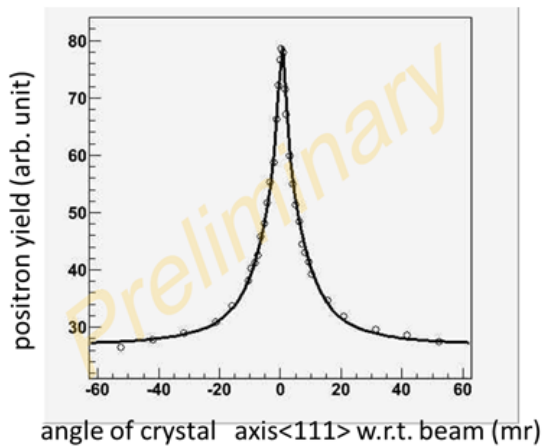

Figure 3. Positron yield as a function of crystal axis with respect to the incident electron beam.

After aligning crystal axis to the electron beam, positron yield were measured systematically for various conditions. Two cases for crystal-amorphous hybrid configuration; crystal was aligned/off-aligned to electron beam, which are referred as hybrid on axis and hybrid off axis here after. Conventional configuration; in which driving electrons are directly impinged onto amorphous targets. For hybrid cases, the sweeping magnet was always excited so that only photons from the crystal were impinged onto the amorphous targets. Positron yields from amorphous targets were measured for positron momentum of 5, 10, and 20 MeV respectively. Typical results for 20 MeV positrons are plotted in Figure 4. Detail of the data is being analyzed. Particularly, the aperture of the detection system, the analyzing magnet and the collimator system, are crucial to understand observed data. A preliminary simulation study showed that number of positron reached at the detector per incident electron is less than 0.1%. It is mainly due to aperture of the collimators. Since it is known that angular distribution of positrons from amorphous targets are different from the hybrid and conventional configuration, detail investigation is necessary to evaluate performance of the hybrid scheme as a positron source from the experimental data.

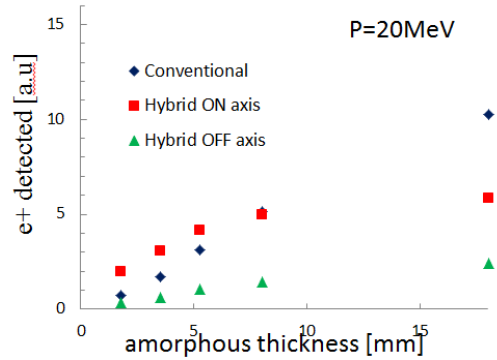

Figure 4. Positron yields with different amorphous target thickness for Conventional, Hybrid ON axis and OFF axis.

## 3.2 Temperature measurement

In order to evaluate energy deposition in the amorphous target, thermocouples are attached on the back of 8 mm and 18 mm amorphous target. With thermocouples, two kind data were recorded. One is temperature reached at the equilibrium after injecting beams onto targets. This temperature is expected to be proportional the total energy deposit in the target. Other data is an instantaneous temperature rise at each bunch injection. It is expected to be useful to extract information for the Peak Energy Deposit Density (PEDD) in the target. The temperature at the equilibrium was plotted for various configurations in Figure 5. The vertical axis of the plot is measured temperature rise from the room temperature. The horizontal axis is the total energy deposit for each configuration calculated by numerical simulation using GEANT4. Proportionality between the temperature rise and total energy deposit was observed except for a data point with hybrid configuration with the amorphous target of 18 mm thick. In this case, the thickness of the amorphous target is thicker than the depth of the shower maximum. Since thermocouples measured the temperature of back of the amorphous target, it does not show the highest temperature inside the target if it was thicker than the depth of the shower maximum. In order to understand data more precisely, estimation of heat flow by a finite element method is planned. Figure 6 shows bunch by bunch temperature rise again measured for various configurations. The vertical axis is the instantaneous temperature rise at each bunch injection while horizontal axis is the PEDD at the back of the amorphous target obtained by the GEANT4 simulation. Proportionality between the temperature rise and PEDD was observed but again detail investigation is in progress to understand relation between the measured temperature and calculated PEDD.

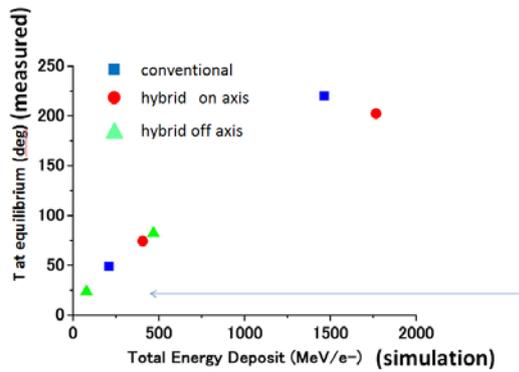

Figure 5. Temperature at the back of the amorphous target t itsa equilibrium versus calculated total energy deposit.

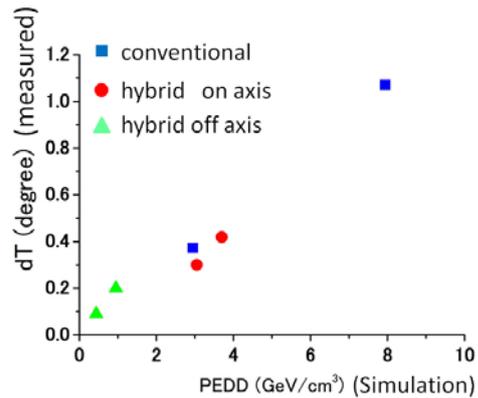

Figure 6. Bunch by bunch temperature rise at the back of the amorphous target versus calculated PEDD.

## 4  Summary

Experimental study on the hybrid target is in progress using the 8 GeV electron beam at KEKB LINAC. The data for positron yield for various conditions were accumulated systematically, by changing thickness of amorphous target, different positron momentum, hybrid/conventional configuration. Temperatures of amorphous targets were also measured to obtain information of energy deposit in the target to estimate difference of heat load on the target for the hybrid and conventional case. Toward the evaluation of the hybrid scheme as positron sources from, we need;
- Numerical simulation of detector system including the aperture of the analyzing magnet and the collimator system.
- Evaluation of heat flow in the target and comparison with measured temperature.

In addition, it was found that temperature of amorphous targets at the equilibrium as well as time dependence of ms time range were possible with thermocouples, we plan to make an array of thermocouples observe two dimensional heat flow on the amorphous target.

## 5  Acknowledgments

The authors would like to thank KEKB accelerator clue to provide electron beams during the experiments. Thay also thank members of ILC/CLIC collaboration on the source R&D for valuable discussion. This work is supported Grant-Aid in Scientific Research in MEXT, Japan (# 22540314) .

## 6  References

[1]  ILC Reference Design Report  ILC-REPORT-2007-001(2007)
[2] CLIC Conceptual Design Report http://clic-study.org/accelerator/CLIC-ConceptDesignRep.php
[3] See for example, X. Artru et. al, Positron sources using channeling: A promising device for linear colliders ,NuovoCimento C2011.6 141-- 148 , 34,4